# Adhesion mechanics of graphene on textured substrates

Narasimha G Boddeti[1], Rong Long[1], and Martin L Dunn[2]

[1]*Department of Mechanical Engineering, University of Colorado, Boulder, Colorado 80309, US*

[2]*Singapore University of Technology and Design, Singapore, 487372*

Graphene, the two dimensional form of carbon, has excellent mechanical, electrical and thermal properties and a variety of potential applications including nano-electro-mechanical systems, protective coatings, transparent electrodes in display devices and biological applications. Adhesion plays a key role in many of these applications. In addition, it has been proposed that the electronic properties of graphene can be affected by elastic deformation caused by adhesion of graphene to its substrate. In light of this, we present here a continuum mechanics based theoretical framework to understand the effect of nano-scale morphology of substrates on adhesion and mechanics of graphene. In the first part, we analyze the adhesion mechanics of graphene on one and two dimensional periodic corrugations. We carried out molecular statics simulations and found the results to be in good agreement with our theory. We modeled adhesive interactions surface forces described by Lennard-Jones 6-12 potential in both our analysis and simulations and in principle can be extended to any other interaction potential. The results show that graphene adheres conformally to substrates with large curvatures. We showed in principal that the theory developed here can be extended to substrates of arbitrary shapes that can be represented by a Fourier series.

In the second part, we study the mechanics of peeling of graphene ribbons from one dimensional sinusoidally textured substrates. In the molecular statics simulations, we observed two key features in the peel mechanics of the ribbons - the ribbons slide over the substrate and undergo adhesion and peeling near the crack front in an oscillatory manner, the frequency of which reveals the wavelength of the underlying substrate. Our theory qualitatively captures these features of the peel mechanics and is general enough that it can be extended to other two dimensional materials like Molybdenum Disulphide (MoS2), Boron Nitride (BN) or other thin films and different kinds of interaction potentials.

## I.     INTRODUCTION AND BACKGROUND

Graphene is a two-dimensional crystalline allotrope of carbon with desirable properties[1,2] like high Young's modulus (~1 TPa) and mechanical strength,[3] low defect density, chemical inertness,[4] and high thermal and electrical conductivity.[5,6] In addition, it is one atomic layer thin and has a bending rigidity comparable to those of biological membranes (~ 1 eV)[7,8] making it a prototypical membrane material. The low bending rigidity allows graphene to be extremely flexible and conform well to the underlying substrates as evidenced in experiments like Lui et al's.[9] Some studies have realized that graphene's electronic properties can be altered in an useful manner using mechanical strain.[10–13] Understanding what makes a graphene membrane conform well or otherwise will help in designing novel electronic devices that will take advantage of the strains that develop as a result of adhesion. In addition, this can aid the design of substrate morphologies to alter the adhesive properties of graphene and other materials which in turn could aid in developing better graphene based protective coatings, transparent electrodes, flexible electronics and nano-electro-mechanical systems.[4,14–17]

In the literature, the effect of substrate morphology on membrane (especially biological/soft membranes) adhesion has been extensively studied in the continuum setting.[18–21] The general strategy is to construct a free energy functional, $F$ which includes the elastic bending and stretching strain energies of the membrane ($F_{ben}$ and $F_{str}$) along with the adhesion energy due to the membrane's interactions with the substrate ($F_{adh}$). If the substrate topography is described by a function $s(x,y)$, then mathematically the goal is to obtain the shape attained by the membrane, $m(x,y)$ so as to minimize the free energy functional, $F$.

$$F(m(x,y)) = F_{ben} + F_{str} + F_{adh}, \qquad (1)$$

$$\begin{aligned}F_{ben} &= \int dA \, \frac{1}{2} D \left((\kappa_x + \kappa_y)^2 - 2(1-\nu)(\kappa_x \kappa_y - \kappa_{xy}^2)\right) \\ F_{str} &= \int dA \, \frac{1}{2} C \left((\epsilon_x + \epsilon_y)^2 - 2(1-\nu)(\epsilon_x \epsilon_y - \epsilon_{xy}^2)\right) \\ F_{adh} &= \int dA_m \int dA_s V_{pot}(s,m).\end{aligned} \qquad (2)$$

Here, $dA$ and $dA_m$ are the area elements on undeformed and deformed membrane respectively, $dA_s$ is the area element on the substrate, $D$ and $C$ are bending and stretching rigidities respectively, $\kappa_\alpha$ and $\epsilon_\alpha$ are the membrane curvature and strain along $\alpha$ ($\alpha = x$, $y$ or $xy$) and $V_{pot}$ is the interaction potential between the atoms of the substrate and the membrane. With any realistic potential function, this is a complicated problem to solve even numerically. Hence the problem is usually reduced, with companion simplifications, to one dimension with a periodic pattern for the substrate like a sine function.

With the advent of 2D crystals like graphene and the ability to examine their morphology accurately using scanning probe techniques like AFM, this problem has been revisited recently in the literature[22–25] with essentially the same continuum approach as described. Each work made the necessary simplifications to arrive at their primary conclusion that the conformity of graphene on a given substrate depends on the substrate morphology, adhesion strength and the number of layers. It has been found that on 1D sinusoidally corrugated substrates, there is a snap through phenomenon where a graphene membrane goes from being non-conformal to conformal as the amplitude or wavelength of the corrugation is changed. Here in this case, conformal is the configuration where the ratio of corrugation amplitudes of graphene membrane and the substrate is close to 1; while non-conformal is the configuration where it is close to zero. This phenomenon has also been observed experimentally.[26] Here in this paper, we pursue this problem to obtain a general understanding of effect of nano-scale roughness on adhesion both analytically as well as numerically. The analytical approach we take here will differ from the existing ones in literature in how the adhesion energy is calculated and we compare our results with those in the literature. Specifically, we use an extended form of Derjaguin approximation to calculate the adhesion energy. The numerical approach will depart from the continuum setup altogether by using 'molecular mechanics/statics' simulations. This also allows us to look at the atomistic details of the adhesion mechanisms of the graphene membranes while validating the continuum model.

We also study the mechanics of peeling on textured substrates by simulating 'peel test' of a graphene ribbon. Here, we quasi-statically delaminate a graphene nano-ribbon by displacing one end perpendicularly to the horizontal plane while pinning the other end. In molecular statics simulations, we observed equilibrium configurations consistent with the ribbons sliding on the substrate. As they slide, the conformity of the ribbon reduces gradually until there is a peel event. After this event, conformity is recovered partially and the ribbon starts to slide again. We noticed that each of the peel events trigger a discontinuity in the magnitude of the peel force and that the peel force has periodicity commensurate with that of the substrates. We developed an approximate theory that captures the essential features of the simulations qualitatively, and reasonably well quantitatively. These results suggest that graphene ribbons might be useful to scan and probe the atomic scale roughness of rigid substrates with the help of the understanding we developed here.

## II. MORPHOLOGY OF GRAPHENE ON TEXTURED SUBSTRATES

A. Theory

To reiterate the problem at hand, we consider a graphene membrane adhered to a rigid substrate as illustrated in Fig. 1. Given the functional form of the substrate surface $s(\vec{\rho})$, the goal is to find the functional form of the graphene membrane $g(\vec{\rho})$ ($\vec{\rho}$ being the position vector) with a given operant interaction potential, $V_{pot}$ between the substrate and graphene at the interface. This potential is assumed to be van der Waals interactions (vdW) between atoms that can be described by a Lennard-Jones (LJ) potential. Starting with LJ 6-12 potential $V_{pot}(r) = 4\epsilon\left(\left(\frac{\sigma}{r}\right)^{12} - \left(\frac{\sigma}{r}\right)^{6}\right)$ ($\epsilon$ is depth of the potential well and $\sigma$ is the distance at which magnitude of the potential is zero), one can then arrive at a continuum expression via direct integration for the potential, $V_f$ that acts between two flat atomic surfaces separated by a distance $h$ (the subscript $f$ is to signify that this is the potential for two flat surfaces):

$$V_f(h) = \rho_A^2 \int_0^\infty 4\epsilon\left(\left(\frac{\sigma}{(r^2+h^2)^{\frac{1}{2}}}\right)^{12} - \left(\frac{\sigma}{(r^2+h^2)^{\frac{1}{2}}}\right)^{6}\right) 2\pi r dr$$
$$= -\gamma_0\left(\frac{5}{3}\left(\frac{h_0}{h}\right)^4 - \frac{2}{3}\left(\frac{h_0}{h}\right)^{10}\right). \quad (3)$$

Here, the integration is done in cylindrical coordinates with the radial coordinate being denoted by $r$ and $\rho_A$ is the areal density of the atoms. It can be easily verified that here $h_0 = \sigma$ is the equilibrium separation where the potential has a minimum and the force between the two flat surfaces is zero. The adhesion energy per unit area, $\gamma_0$ is related to other terms via $\gamma_0 = 6\pi\rho_A^2\sigma^2\epsilon/5$.

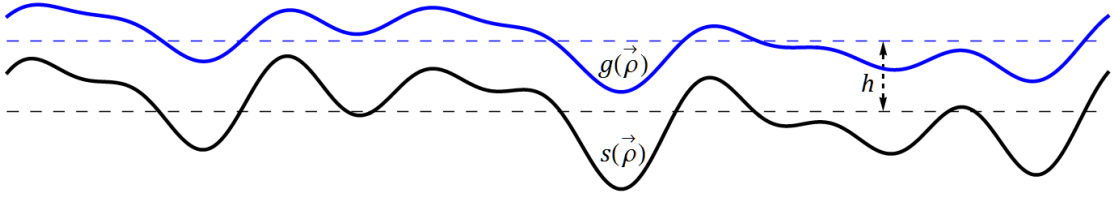

*Figure 1 (Color Online) Illustration showing the morphology of a graphene membrane (blue) on a corrugated substrate (black) with the dashed lines depicting the mean height*

For two arbitrarily shaped surfaces such as the ones shown in Fig. 1, the vdW potential is nonlocal i.e. it depends on the functional forms of the interacting surfaces ($V_{pot} = V_{pot}(g(\vec{\rho}), s(\vec{\rho}))$) and as mentioned before, is difficult to calculate even numerically. Hence, we borrowed and extended the approach used by Palasantzas and Backx[19] and Swain and Andelman[20] where they used Derjaguin approximation to simplify the problem. The Derjaguin approximation expresses the energy between two surfaces or bodies due to an interaction like vdW attraction, $V_{pot}$ as a function of the local separation only. In mathematical terms:

$$V_{pot}(g(\vec{\rho}), s(\vec{\rho})) \approx V_f(g(\vec{\rho}) - s(\vec{\rho})). \tag{4}$$

Now, we can write the free energy of the system, $F(g(\vec{\rho}))$ as:

$$F(g(\vec{\rho})) = F_{ben} + F_{adh} \\ = \int dA \frac{D}{2}\left((\kappa_x + \kappa_y)^2 - 2(1-\nu)(\kappa_x \kappa_y - \kappa_{xy}^2)\right) + \int dA\, V_f(g(\vec{\rho}) - s(\vec{\rho})). \tag{5}$$

Here the contribution due to stretching is neglected completely as it is assumed that the interfacial friction is very small and the graphene membrane should be able to slide on the substrate freely. The Derjaguin approximation in effect replaces the surfaces with a series of parallel flat plates and calculates the total adhesion energy by adding the interaction potentials between these sets of parallel plates. Thus, Derjaguin approximation works best when the surfaces involved have small slopes. Even with these simplifications the potential is still not tractable to solve for $g(\vec{\rho})$. Palasantzas and Backx[19] and Swain and Andelman[20] expanded the integrand in the second integral about the equilibrium separation $h_0$ to the second order:

$$V_f\big(g(\vec{\rho}) - s(\vec{\rho})\big) = V_f(h_0) + \frac{d^2 V_f(h)}{dh^2}\bigg|_{h=h_0} \frac{(g - h_0 - s)^2}{2}. \tag{6}$$

Here they assumed that the mean height of the substrate is zero and that of the membrane is $h_0$ and that $g - h_0 - s \ll 1$. We extend this further by expanding the potential about yet to be determined equilibrium separation $h$ to an arbitrary number of terms, $p$ ($z(\vec{\rho}) = g(\vec{\rho}) - h$):

$$V_f\big(h + z(\vec{\rho}) - s(\vec{\rho})\big) = V_f(h) + \sum_{i=1}^{p} \frac{d^i V_f(h)}{dh^i} \frac{(z - s)^i}{i!}. \tag{7}$$

Swain and Andelman[20] using eq. (6) showed that sinusoidal substrates allow for sinusoidal membrane profiles and a one-to-one correspondence does not hold good for any other arbitrary functions. We assumed that this still holds good here and it can in principal be shown to be the case numerically with the help of the analysis we developed further ahead in this section. Sinusoidal surfaces, though a poor representation of randomly rough surfaces, allow us to simplify the analysis while qualitatively capturing the main features of the underlying physics. We first dealt with one dimensional sinusoidal surfaces i.e. $s(x) = c\,\text{Sin}[q\,x]$ where $c$ and $q$ are the amplitude and wave numbers of the sinusoid respectively. To narrow the search for a minimized free energy state, we assume $z(x) = a\,\text{Sin}[q\,x]$, the free energy per unit area, $\hat{F}$ is:

$$\begin{aligned}
\hat{F}(a,h) &= \int_0^\lambda \frac{dx}{\lambda} \frac{D}{2} \left(\frac{d^2 g}{dx^2}\right)^2 + \int_0^\lambda \frac{dx}{\lambda} V_f(h + z - s) \\
&= \frac{D}{4} a^2 q^4 + V_f(h) + \sum_{i=1}^{\lfloor p/2 \rfloor} \frac{d^{2i} V_f(h)}{dh^{2i}} \frac{(a-c)^{2i}}{4^i (i!)^2}.
\end{aligned} \tag{8}$$

Here $\lambda = 2\pi/q$ is the wavelength and $\lfloor p \rfloor$ is the largest integer not greater than $p$. It is fairly straightforward to formulate the free energy as shown here using a computer algebra system (CAS) and then numerical optimization gives us the equilibrium configuration of the membrane for any arbitrary one dimensional sinusoidal corrugation.

This method can easily be extended to 2D sinusoidal substrates such as an egg crate pattern: $s(x, y) = c \, \text{Sin}[q_x x] \, \text{Sin}[q_y y]$ where $q_x$ and $q_y$ are the wave numbers in x and y directions respectively. The free energy in this case, assuming the membrane will follow $g(x, y) = h + a \, \text{Sin}[q_x x] \, \text{Sin}[q_y y]$, will then be:

$$\hat{F}(a, h) = \frac{D}{8}\left(q_x^2 + q_y^2\right)^2 a^2 + V_f(h) + \sum_{i=1}^{\lfloor \frac{p}{2} \rfloor} \frac{d^{2i} V_f(h)}{dh^{2i}} \frac{(2i)!}{16^i (i!)^4} (a - c)^i. \tag{9}$$

This looks very similar to the free energy expression in eq. (8) for 1D sinusoidal corrugations. Hence again by optimizing the free energy numerically to find $a$ and $h$, we should be able to arrive at the equilibrium configuration of graphene membranes.

This approach can be generalized to work with a full or truncated Fourier series that involves multiple sine or cosine waves of different amplitudes and wavelengths in superposition. If we assume the substrate is represented by the function $s(x) = \sum_{n=-N}^{N} c_n e^{i q_n x}$ ($0 \leq x \leq L$) ($n \in \mathbb{Z}, n \neq 0, q_n = 2\pi n / L$) with complex conjugate coefficients ($c_n = c_{-n}^*$) making it a real valued truncated Fourier series with 2N terms. Then the graphene membrane can be assumed to take the form $g(x) = h + \sum_{n=-N}^{N} a_n e^{i q_n x}$ ($a_n = a_{-n}^*$). The free energy per unit area in this case will be:

$$\hat{F}(\{a_n\}, h) = \frac{D}{2} \sum_{n=-N}^{N} |a_n|^2 q_n^4 + V_f(h)$$
$$+ \sum_{j=1}^{p} \frac{d^j V_f(h)}{dh^j} \frac{1}{j!} \sum_{\substack{\sum_n l_n = j \\ \sum_n n l_n = 0}} \frac{j!}{\prod_n l_n!} \prod_n (a_n - c_n)^{l_n}. \qquad (10)$$

Here, $c_n$ and $a_n$ are the Fourier coefficients and are complex numbers; $h$ (equilibrium separation) and array of coefficients $\{a_n\}$ being the unknowns. The internal summation in the nested summation of the last term is a result of a multinomial expansion where $l_n$ are the set of exponents which have to obey the constraints $\sum_n l_n = j$ and $\sum_n n l_n = 0$ (note that $n$ can take either positive or negative integer values). The second of the constraints comes from the non-zero terms after integration of each term in the multinomial expansion. Also following this approach, similar expression for the free energy can be arrived at for two dimensional substrates represented by a full or truncated 2D complex Fourier series (see Appendix B). With $2N$ terms in the truncated Fourier series, the free energy has $2N+1$ unknowns which can be found as before by optimizing the free energy.

Using the free energy expression in eq. (10), one can show that the solution for the simplest case, $s(x) = c \sin[qx]$ is indeed $g(x) = h + a \sin[qx]$. The amplitudes of the higher frequency components would just turn out to be negligibly small compared to $a$, the amplitude of the lowest frequency component. Later we will use these free energies in eqs. (8)-(10) to numerically calculate the equilibrium membrane profiles for different corrugated substrate profiles and compare the results with molecular statics simulations (see Appendix A for comparison with other methods in the literature[22,27,28]).

B. Simulations – 1D Sinusoidal Corrugations

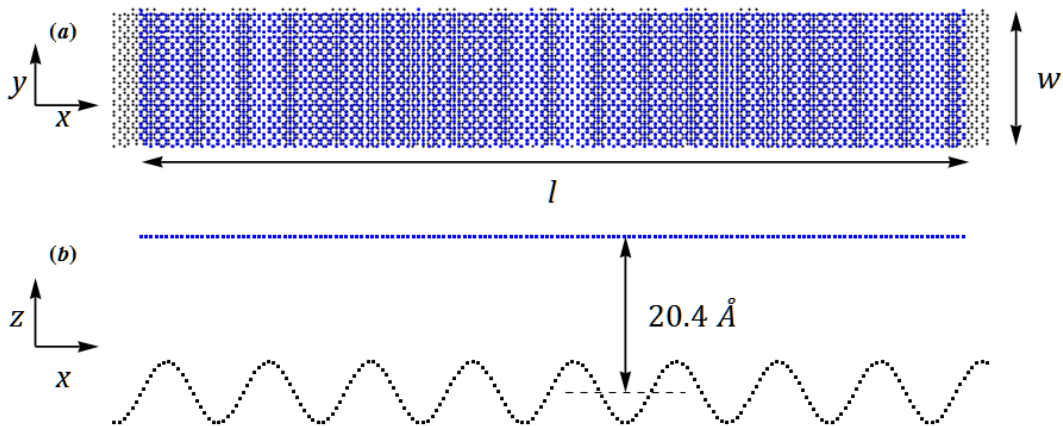

*Figure 2 (Color Online) The initial configuration of the atoms (blue - graphene, black - substrate): (a) Top view (b) Side view with $c = 4$ Å and $\lambda = 24$ Å.*

We carried out molecular statics simulations using LAMMPS[29] initially with 1D sinusoidally corrugated rigid substrates where we varied the amplitude ($c$) and wavelength ($\lambda$) of the substrates in a systematic manner to determine the effect on the graphene membrane conformity. The simulation setup consists of a fictitious graphene-like substrate with just a single layer of atoms. The substrate atoms are pre-arranged in a sinusoidal manner with the desired amplitude and wavelength. The atoms in the graphene membrane interact via AIREBO potential (Adaptive Interatomic Reactive Empirical Bond Order)[30] which accounts for covalent bonding at short distances (~2 Å) and vdW interactions at larger distances (> 3 Å) through a prescribed LJ 12-6 potential with a cut-off distance of 10.2 Å (the cut-off distance is the distance beyond which the interaction is zeroed). The whole initial setup is as shown in Fig. 2 with black colored dots denoting the substrate atoms and blue colored dots denoting the graphene atoms. Initially, the graphene atoms in a flat configuration are vertically set apart by 20.4 Å from the substrate well beyond the LJ cutoff distance so that there are no vdW interactions. Periodic boundary condition is applied along the width direction while the graphene atoms are free to move in the length direction. The substrate is made slightly longer than the graphene membrane to accommodate vdW interactions near the graphene membrane edges. The reason for choosing to represent the substrate with just one layer of atoms is primarily that it saves computational effort. It is also easier to setup in comparison to a substrate with bulk atoms and should be able to capture the essential physics even without any bulk atoms.

After setting up the atoms, the interaction potentials and the boundary conditions, the graphene atoms are allowed to relax while the substrate atoms are fixed. At the end of this minimization step, the graphene atoms in their relaxed configuration are moved closer to the substrate atoms by about 13 Å from the initial mean separation of 20.4 Å. The graphene atoms, now within LJ potential cut-off, start to interact with the substrate atoms while the substrate atoms are still rigidly fixed. Under the influence of these interactions, in what will be the second energy minimization step, the graphene membrane moves closer to the substrate until an equilibrium configuration is reached. The difference between the total energies at the end of the second and the first minimization steps gives the apparent adhesion energy; dividing it by the area of the graphene sheet gives apparent adhesion energy per unit area, $\gamma$. This is because at the end of the first minimization step, the graphene atoms are in a relaxed flat configuration and are not interacting with the substrate atoms; while at the end of the second minimization step the atoms are deformed and adhered to the substrate. Hence, the difference of energies of these two configurations gives us the apparent adhesion energy which in turn is the energy gained by the system due to adhesive interactions between the substrate and graphene atoms and the energy lost due to bending of the graphene atomic bonds.

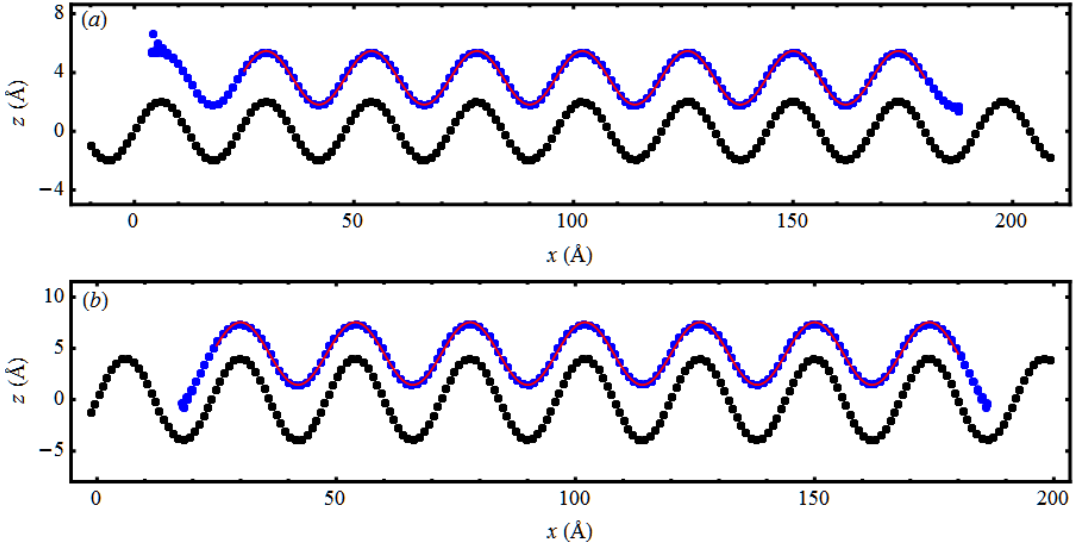

*Figure 3 (Color Online) The final equilibrium configurations for (a) $c = 2$ Å and (b) $c = 4$ Å with $l = 194$ Å, $w = 50$ Å, $\lambda = 24$ Å and $\gamma_0 = 0.3$ J/m2. The blue and black dots are atoms in graphene and the substrate respectively while the red curve is the fitted sine curve.*

| $w$ (Å) | $c = 2$ Å | | | $c = 4$ Å | | |
|---|---|---|---|---|---|---|
| | $a$ (Å) | $h$ (Å) | $\gamma$ (J/m$^2$) | $a$ (Å) | $h$ (Å) | $\gamma$ (J/m$^2$) |
| 49.59 | 1.8249 | 3.6360 | 0.2656 | 3.0464 | 4.4910 | 0.1888 |
| 62.15 | 1.8249 | 3.6360 | 0.2641 | 3.0462 | 4.4911 | 0.1878 |
| 74.71 | 1.8262 | 3.6356 | 0.2633 | 3.0446 | 4.4918 | 0.1875 |

*Table I The results of the simulations with varying widths for the graphene membrane with the length fixed at about $l = 194$ Å, $\lambda = 24$ Å and $\gamma_0 = 0.3$ J/m$^2$*

All the simulations are performed at a temperature of 0 K, any effects of finite temperature are not considered here. We used conjugate gradient method for all the minimization steps. The initial set of simulations are performed with $\lambda = 24$ Å, $c = 2$ and 4 Å and for monolayer graphene while varying the length ($l$) and the width ($w$) of the graphene membrane. This exercise is done to make sure that the results are not sensitive to the size of the system and any edge effects due to the finite size of graphene are negligible. First, the width of the graphene sheet is varied from about 50 Å to 62 Å to 75 Å while keeping the length fixed at about 194 Å.

For each simulation, assuming the graphene membrane takes the form $g(x) = h + a \sin[\frac{2\pi}{\lambda}x]$, the amplitude ($a$) and mean separation ($h$) for the graphene membrane are extracted via curve fitting (details in the supplementary text) from the final equilibrium configuration along with the effective adhesion energy per unit area ($\gamma$). The results of these simulations are shown partially in Fig. 3 and tabulated completely in Table Table *I*. The figure shows that the graphene atoms (blue dots) follow a sine curve (in red) very closely. From the table, it is clear that we get about the same result in each case even with fewer atoms when the width is about 50 Å.

| $l$ (Å) | $c = 2$ Å | | | $c = 4$ Å | | |
|---|---|---|---|---|---|---|
| | $a$ (Å) | $h$ (Å) | $\gamma$ (J/m²) | $a$ (Å) | $h$ (Å) | $\gamma$ (J/m²) |
| 193.7 | 1.8249 | 3.6360 | 0.2656 | 3.0464 | 4.4910 | 0.1888 |
| 290.4 | 1.8021 | 3.6398 | 0.2641 | 2.2880 | 4.9947 | 0.1643 |
| 387.1 | 1.8085 | 3.6367 | 0.2597 | 1.9052 | 5.2855 | 0.1497 |

*Table II The results of the simulations with varying lengths for the graphene membrane with the width fixed at about w = 50 Å, λ = 24 Å and $\gamma_0$ = 0.3 J/m²*

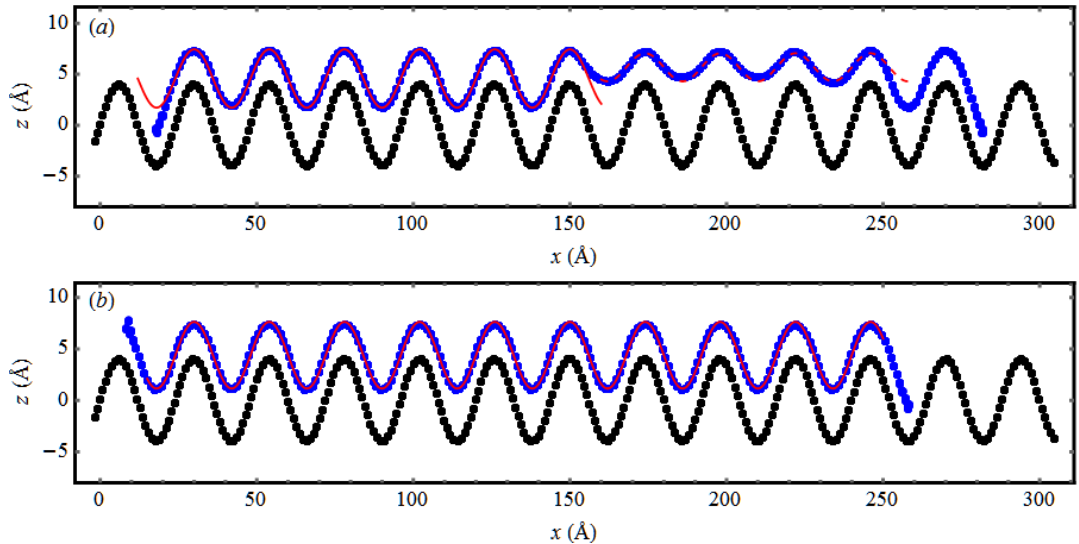

*Figure 4 (Color Online) The final equilibrium configurations for a graphene membrane of l ≈ 290 Å with c = 4 Å, w = 50 Å, λ = 24 Å and $\gamma_0$ = 0.3 J/m² – (a) with flat initial configuration, (b) with sinusoidal initial configuration. The blue and black dots are*

*atoms in graphene and the substrate respectively while the red curves are the fitted sine curves.*

Similar study is done with varying lengths for the graphene membrane while keeping the width fixed at about 50 Å. The results are shown partially in Fig. 4 and tabulated completely in Table Table *II*. We can see that with $c = 2$ Å, the results are practically the same with different lengths; however with $c = 4$ Å, the results differ with the shorter graphene membranes conforming better than the longer ones. This is probably due to the inability of the energy minimization step to reach the absolute minimum. The graphene membrane reaches what might be an intermediate equilibrium configuration where the conformity is not quite uniform as seen in Fig. 4(a) for $l = 290$ Å case and the fitting procedure used gives poor results as shown in Table Table *II*. The figure also shows the sine curve fitting done to the two different regions of the membrane in red and the solid curve is closer to the result obtained with shorter graphene membrane. To ascertain which of these two fitted sine curves with $(a, h) = (2.86, 4.59)$ Å and $(1.38, 5.69)$ Å corresponds to the actual minimum, we repeated the simulation with the graphene atoms initially along one of the aforementioned fitted sinusoidal curves instead of a flat shape. The simulations in each case produced results of $(3.23, 4.39)$ Å and $(3.11, 4.45)$ Å for the initial configurations of $(2.86, 4.59)$ Å and $(1.38, 5.69)$ Å respectively. These results are in turn are close to the one obtained for the shorter membrane i.e. $(3.05, 4.49)$ Å.

In view of the above discussion and results shown in Tables Table *I* and Table *II*, we concluded that we get about the same results with different lengths and widths for graphene. So we used $l \approx 194$ Å and $w \approx 50$ Å for the rest of our simulations knowing that we lose very little in terms of accuracy. Now, we varied the wavelength, $\lambda$ of the substrate from 12 Å to 36 Å in steps of 6 Å while keeping the amplitude, $c$ fixed at either 1 or 2 Å. We also solve for the equilibrium configuration in each case using our theory where we used 1 eV for monolayer graphene membranes bending rigidity in line with the values found in the literature[7,8,31]. We also obtain a similar value using molecular statics simulations, the details of which are available in the supplementary text. The results of these simulations (red dots) are shown in Fig. 5 along with the theoretical calculations (black curves). We see good agreement between the simulations and the theory in general. With increasing wavelength the conformity of the graphene membrane to the substrate changes from poor to good leading to increase in the membrane amplitude, decrease in the equilibrium separation and increase in the adhesion energy.

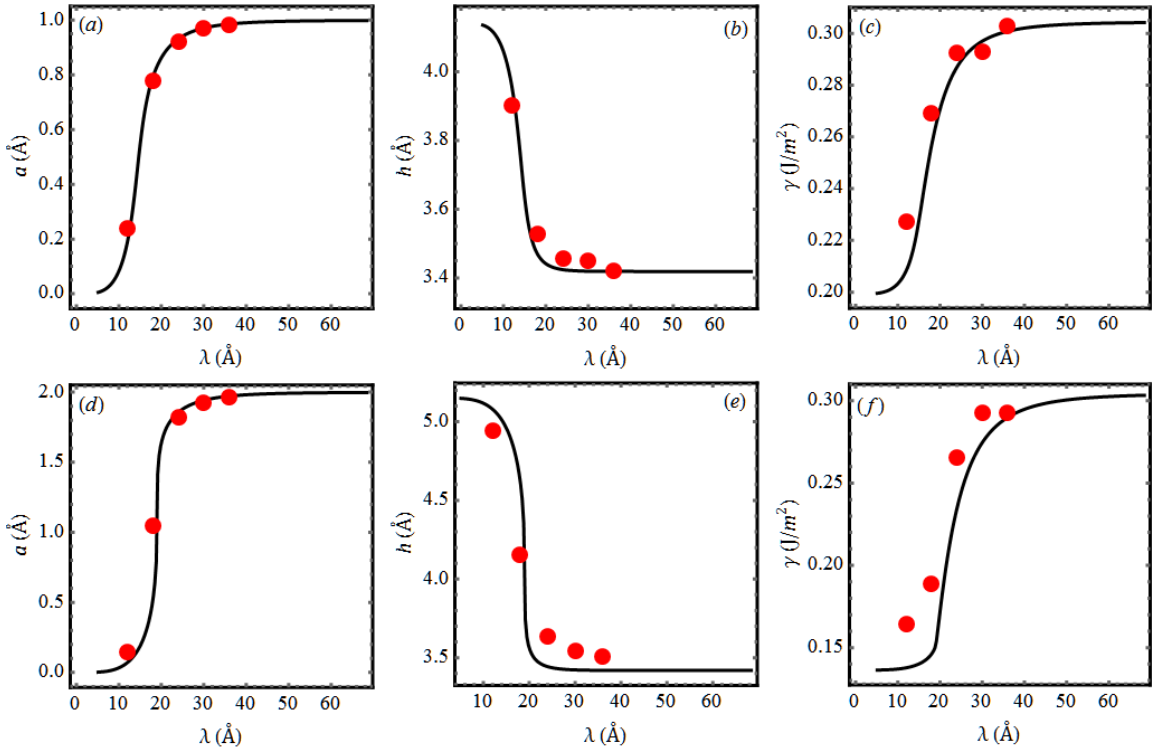

*Figure 5 (Color Online) Plots showing the variation of equilibrium (a, d) amplitude, $a$, (b, e) separation, $h$ and (c, f) adhesion energy, $\gamma$ with the substrate wavelength, $\lambda$. The black curve is from our theory and the red dots are from the simulations. The top (a, b, c) and the bottom rows (d, e, f) show results for substrate amplitude $c = 1$ Å and 2 Å respectively.*

We also carried out simulations where we varied the substrate amplitude, $c$ while fixing the wavelength, $\lambda$ and using the same values for $l$, $w$ and $\gamma_0$ as before. The results are shown in

Fig. 6. Again it can be seen that the theory performs reasonably well at predicting the equilibrium configurations of the graphene membrane as well as the transition from good to poor conformity. The discrepancy between the theory and simulation at higher amplitudes ($c > 3$ Å) might be a result of the limitation of the Derjaguin approximation we used which does a poor job when the curvatures of the surfaces involved are large.

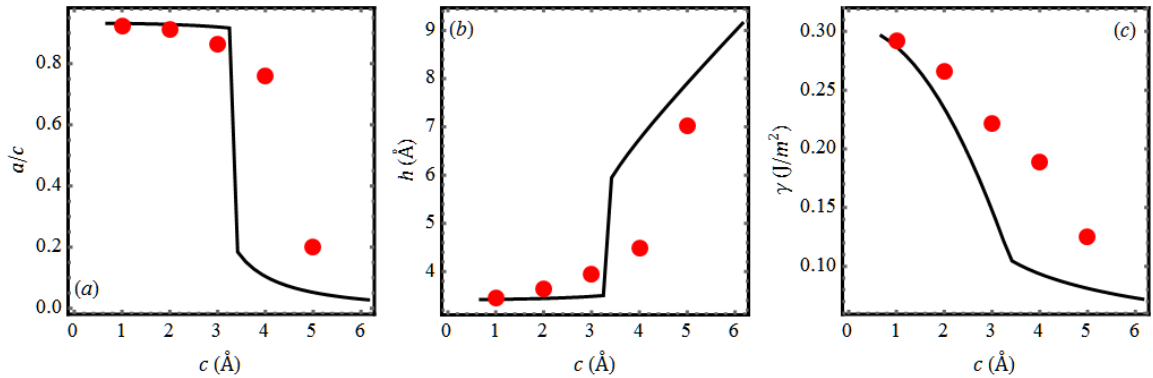

*Figure 6 (Color Online) Plots showing the variation of equilibrium (a) amplitude, $a$ (normalized with respect to substrate amplitude, $c$), (b) separation, $h$ and (c) adhesion energy, $\gamma$ with the substrate amplitude, $c$. The black curve is from our theory and the red dots are from the simulations. For these simulations, the substrate wavelength is fixed at $\lambda = 24$ Å.*

In essence, as we increase the curvature of the substrate, either by increasing $c$ or decreasing $\lambda$, the conformity of the graphene membrane decreases. This is due to the competition between the adhesion and the bending strain. While the adhesive interactions pull the membrane towards the substrate, the bending strain prevents the membrane from completely conforming to the corrugated substrate. The final equilibrium configuration is attained as a balance between these two opposing tendencies is reached. At smaller wavelengths or larger amplitudes, the bending strain is too high leading to poor conformity and at higher wavelengths or smaller amplitudes, bending strain is small enough for the membrane to achieve high conformity.

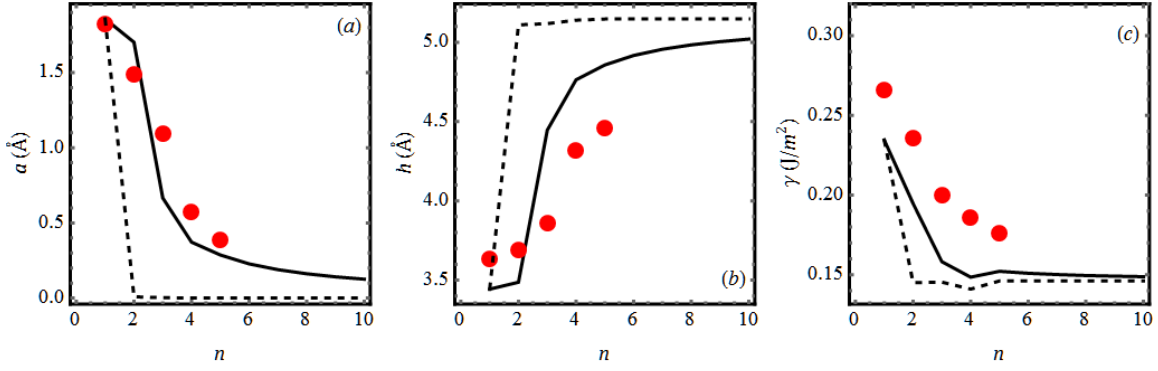

*Figure 7 (Color Online) Plots showing the variation of equilibrium (a) amplitude, $a$, (b) separation, $h$ and (c,f) adhesion energy, $\gamma$ with the number of layers, $n$. The black curves are from our theoretical calculations with different bending rigidities and the red dots are from the simulations. For these simulations, the substrate amplitude and wavelength are fixed at $c = 2$ Å and $\lambda = 24$ Å while $\gamma_0 = 0.3$ J/m².*

Next, we varied the number of layers from 1 to 5 in the graphene membranes. For these set of simulations we fixed the amplitude and wavelength of the substrate at 2 Å and 24 Å respectively while the same parameters are used for LJ potential. The results are shown in Fig. 7. As before, the red dots are from the simulations and each of the two black curves are from our theory with the bending rigidities calculated in two different ways. The dashed curve is obtained using the formula $\kappa_n = nk_1 + Es^3(n^3 - n)/12$ where $n$ is the number of layers, $\kappa_n$ is the bending rigidity of $n$ layer graphene, $E$ is the Young's modulus and $s$ is the inter-layer separation in multi-layered graphene. This relation is obtained from numerical calculations of spherical graphene using "revised periodic boundary conditions" in density functional tight binding method based simulations.[32] We used a value of 1 eV for $\kappa_1$ as before. The solid curve in contrast is obtained by simply assuming $\kappa_n = n\kappa_1$, which meant that each layer in multi-layered membranes behaved independently. As can be seen, we get a better agreement with the simulations with the case where the bending rigidity is assumed to vary linearly. The first approach to calculating the bending rigidity is closer to the straightforward continuum mechanics approach where bending rigidity is simply given by $\kappa = Et^3/12(1 - v^2)$ ($t$ is the thickness);

while the second approach suggests frictionless sliding between layers which seems to be the case in the simulations.

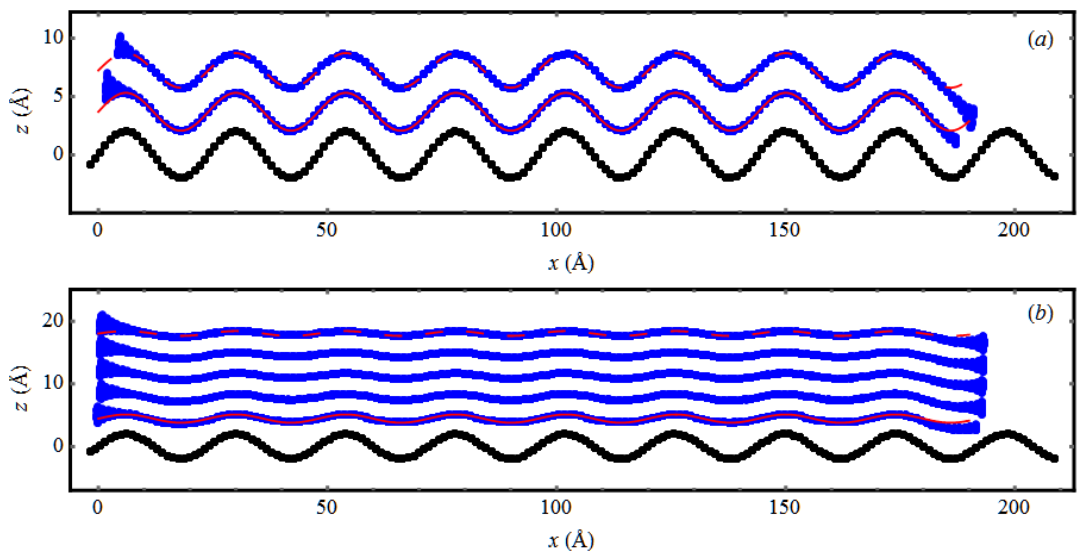

*Figure 8 (Color Online) Plots showing the equilibrium configuration of (a) bilayer and (b) five layered graphene. The black and blue dots denote substrate and graphene atoms while the red curves denote the best fit sine curves for each layer.*

It is also to be noticed that we assumed here implicitly that all the layers will have the same amplitude but this is not the case in the simulations. It is observed in the simulations that the amplitude of each layer decreases progressively from the bottom to the top layers, bottom being the closest to the substrate. To illustrate this point, the equilibrium configurations for two and five layered membranes are shown in Fig. 8. For the bilayered membrane the amplitudes of the bottom and top layers are 1.62 Å and 1.48 Å respectively; while the same for five layered membrane are 0.65 Å and 0.38 Å. This behavior can easily be explained by the nature of the LJ potential. The LJ potential energy decreases rapidly as separation is increased from the equilibrium value, hence the bottom most layer interacts in the strongest manner with the substrate. In fact, due to the cutoff distance for LJ interactions in the simulations there is zero interaction between the substrate and any layer or atoms beyond 10.2 Å. Hence the top layers interact appreciably only with their neighboring layers. In addition to this, the low shear modulus of graphene allows the graphene layers to slide and accommodates varying degrees of bending strain. The net effect of these conditions is that the top layers only react to the corrugations of the layer below and so on leading to progressively decreasing amplitudes from the bottom to the top layers or vice versa. This in turn leads to smoother topographies and decreased adhesion energies for multi-layered graphene membranes when compared to monolayer graphene as evidenced in our simulation results.

C. Simulations – 2D Sinusoidal Corrugations

Having performed simulations with corrugations in one direction alone, we directed our attention to corrugations in both $x$ and $y$ directions. The first step of these simulations involved preparation of substrate. This is achieved by moving atoms on a flat surface out of plane according to the equation $s(x,y) = c \, \text{Sin}\left[\frac{2\pi x}{\lambda_x}\right] \text{Sin}\left[\frac{2\pi y}{\lambda_y}\right]$ so that it forms a structure that looks like an egg crate. This structure is allowed to relax as much as possible by constraining the atoms to move only along the surface given by $s(x,y)$. The graphene membrane, which is initially placed 20.4 Å away from the substrate, is then brought closer and allowed to move to an equilibrium configuration through energy minimization while the substrate atoms are rigidly fixed. The initial and the final equilibrium configurations are as shown in Fig. 9. The equilibrium configuration of the membrane follows the sinusoidal shape of the substrate and hence can be fit to a sinusoidal surface with the same wavelengths but different amplitude. The same post-processing steps as before are done to obtain the γ, $a$ and $h$. For convenience, we chose $\lambda_x = \lambda_y$. As before, we varied the amplitude, $c$ and the wavelength $\lambda$ of the substrate to see the resultant effect on the graphene membrane conformity. When varying the wavelength, the amplitude is fixed at 1 Å and while varying the amplitude, the wavelength is fixed at 24 Å. All the simulations are carried out at 0 K and with monolayer graphene of size 190×190 Å while the substrate is slightly larger to accommodate vdW interactions of the atoms along the edges of the graphene membrane.

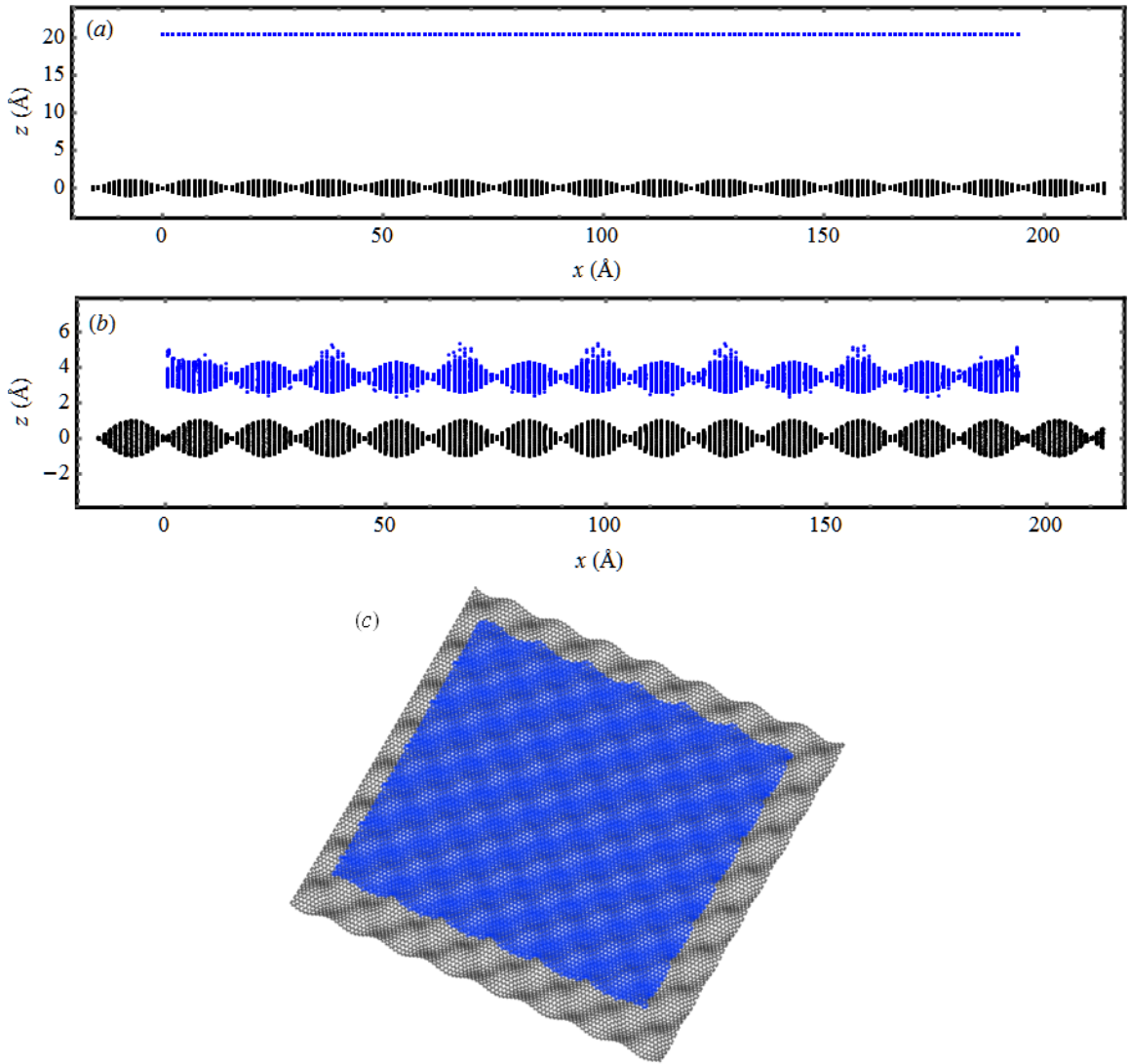

*Figure 9 (Color Online) Side views of (a) the initial system configuration at the beginning of the simulation, and (b) the equilibrium configuration for graphene obtained at the end of the simulation with a substrate amplitude of 1 Å and wavelength 30 Å. General view of the system equilibrium configuration is seen in (c). The black and blue dots denote the atoms in substrate and graphene respectively.*

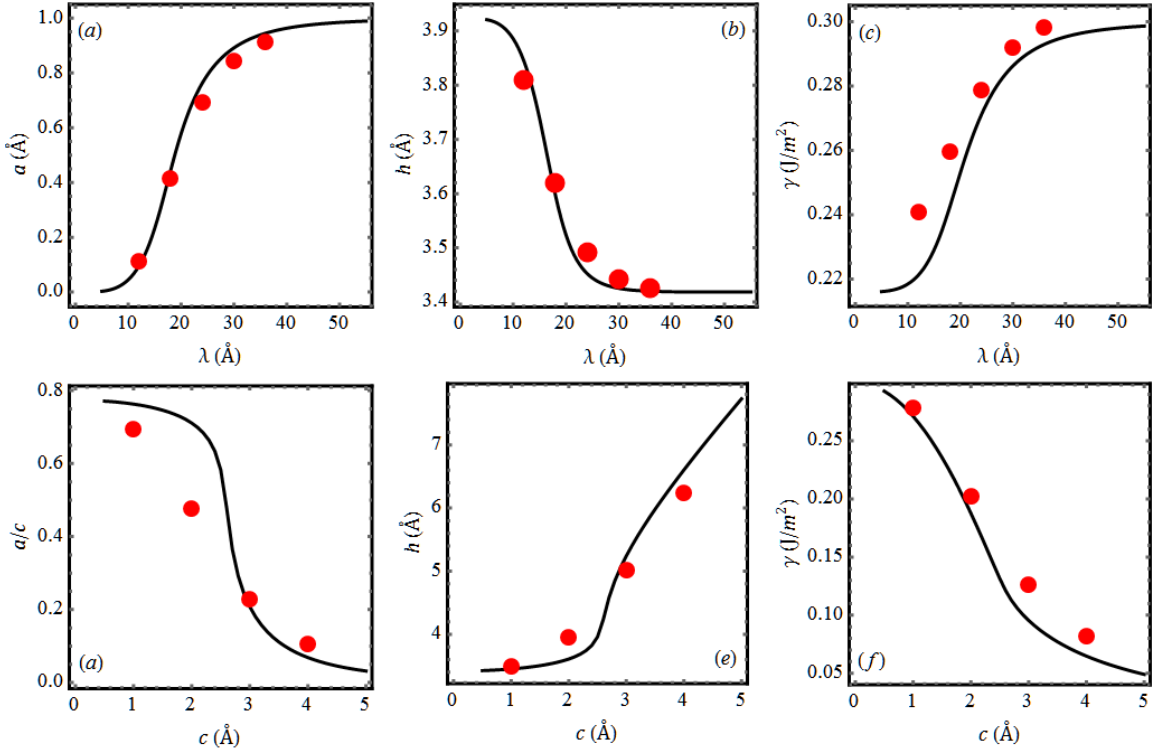

*Figure 10 (Color Online) Plots showing the variation of (a) Amplitude, a, (b) Mean separation, h, and (c) Adhesion energy, γ with wavelength λ of the corrugated substrate with amplitude fixed at c = 1 Å. The bottom row plots show the variation of (d) normalized amplitude, a/c, (e) h and (f) γ with respect to substrate amplitude variation with λ = 24 Å. The graphene sheet size is set at about 190×190 Å.*

The results of the simulations with varying wavelength are shown in Figs. 10(a), (b), and (c) and those with varying amplitude are shown in Figs. 10(d), (e) and (f). The red dots in each plot are the results of the simulations while the black curves are obtained from the theory i.e. optimizing the free energy in eq. (9). It can be seen that, just as in the case of one dimensional sinusoidal corrugations, the conformity of the graphene membrane transitions from good to poor with increasing substrate amplitude or decreasing wavelength. However, this transition is more gradual compared to the one dimensional case. Also to be noticed is the good agreement between the theory and the simulation results.

D. Simulations – 1D Multi-component Sinusoidal Corrugations

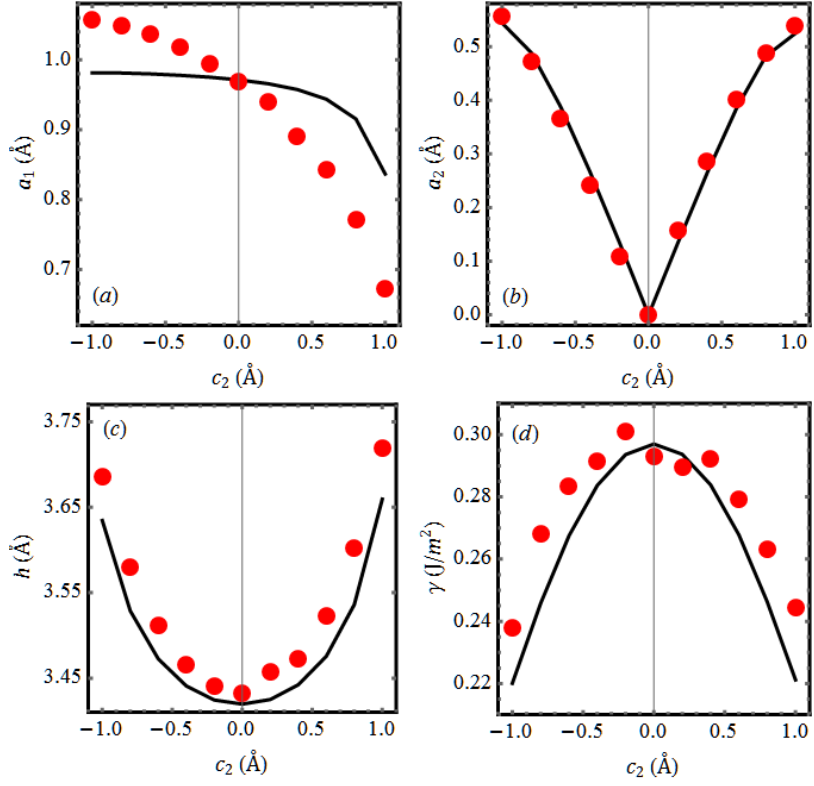

*Figure 11 (Color Online) Plots in the top row show the variation of the amplitudes of each frequency components in the membrane with respect to the amplitude of the higher frequency component in the substrate: (a) $a_1$ vs $c_2$ and (b) $a_2$ vs $c_2$. The plots in bottom row show the variation of the equilibrium separation, h and adhesion energy, $\gamma$ with $c_2$ respectively. The results from the simulations are plotted as red dots and those from simulations are plotted as black curves.*

In section II-A, we discussed a generalized free energy for a substrate described by a full or truncated Fourier series. To retain simplicity, we limited our studies to just two frequency components. We performed simulations with substrates taking the form, $s(x) = c_1 \text{Sin}[qx] + c_2 \text{Cos}[2qx]$ ($q = 2\pi/\lambda$) where we fixed both $c_1$ and $\lambda$ at 1 Å and 30 Å respectively. The simulation setup is exactly same as the one described before with 1D sinusoidal corrugation simulations. The value of $c_2$ is varied from -1 to 1 Å in steps of 0.2 Å and the results are compared with the theory. The free energy for this particular case is given by:

$$\hat{F}(a_1, a_2, h) = \frac{D}{2}\left(\frac{q^4 a_1^2}{2} + \frac{(2q)^4 a_2^2}{2}\right) + V_f(h) \tag{11}$$

$$+ \sum_{j=1}^{p} \frac{d^j V_f(h)}{dh^j} \sum_{k=0}^{j} \frac{1}{k!(j-k)!}(a_1 - c_1)^k (a_2 - c_2)^{j-k} \int_0^{2\pi} \text{Sin}[x]^k \text{Cos}[2x]^{j-k} \frac{dx}{2\pi}.$$

This expression is derived from eq. (10) where the integrand in the last term is expanded with the help of the binomial theorem. The integrals in the last term here can be evaluated analytically for any arbitrary positive integers $j$ and $k$. The free energy in this case has only three unknowns – the amplitude of the lower frequency sine component, $a_1$, the amplitude of the higher frequency cosine component, $a_2$ and the equilibrium separation, $h$. These values are obtained via optimization of $\hat{F}$ as before (here we used $p = 80$). These values along with the adhesion energy, $\gamma$ are also obtained from the simulations using the same post-processing steps as before. Figure 11 shows the simulation results along with those from the analysis. We plot the variation of $a_1$, $a_2$, $h$ and $\gamma$ with respect to $c_2$. It can be seen that the analysis captures the general trend quite well and predicts the amplitude of the higher frequency quite well. As the magnitude of $c_2$ is increased, the overall amplitude of the corrugation is also increased thus decreasing the ability of graphene to conform well. This is reflected quite well in the decrease of adhesion energy and increase of mean separation with increasing magnitude of $c_2$. However, the amplitude of the lower frequency component, $a_1$ shows a curious asymmetric trend. It decreases continuously with increasing $c_2$ and also the analysis does poorly in predicting $a_1$. Looking at the two extreme cases i.e. $c_2 = 1$ and -1 Å as shown in Fig. 12, it is apparent that the two cases are reflections of each other about $z = 0$ albeit out of phase by half a wavelength. However, the membrane prefers to conform closer in the -1 Å case even as there is no appreciable gain in the adhesion energy. This might be explained thus: in each case the higher frequency term can be viewed to be flattening either the peak or valley of the lower frequency sine curve depending on the sign of its amplitude. When positive, it reinforces the valleys and somewhat flattens the peaks and vice-versa while negative. The flattened region is more readily accessible in the positive case than in the negative case leading to the asymmetric behavior. Also, this asymmetric behavior suggests that the adhesion mechanics of substrate surfaces possessing similar statistical properties (amplitude and wavelength) might lead to different scenarios.

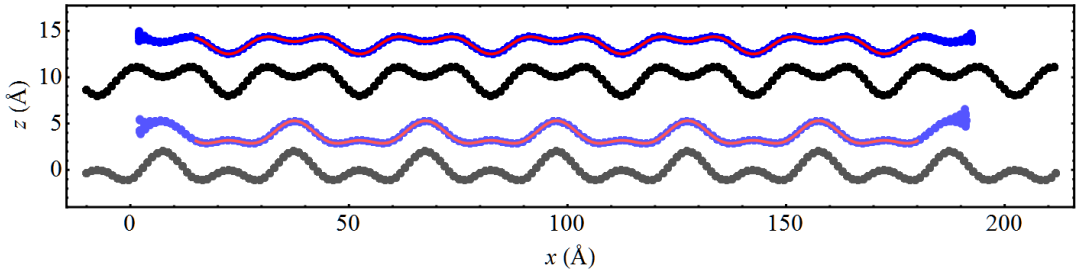

*Figure 12 (Color Online) Equilibrium configurations of graphene membrane obtained from simulations on substrates of the form $s(x) = c_1 Sin[qx] + c_2 Cos[qx]$, with $c_1 = 1$ Å, $q \approx 0.2094$ Å$^{-1}$ and two different values of $c_2 = 1$ Å (above) and -1 Å (below). The blue and black dots are graphene and substrate atoms respectively while the red curves are the fitted curves of the form, $h + a_1 Sin[qx] + a_2 Cos[qx]$.*

Thus we demonstrated here how our analysis can be extended to multi-component corrugations with a dual component 1D substrate profile. The analysis does well, qualitatively at the least, in predicting the conformity and adhesion energy. We surmise that this method might be used to study adhesion qualitatively on simple substrate profiles like square waves and triangle waves since such simple substrate profiles can easily be represented using the few dominant Fourier components in their Fourier series expansions. Such surface profiles can also be readily fabricated to carefully understand and engineer adhesion.

## III. Peeling of Graphene Ribbons on Flat and Corrugated Substrates

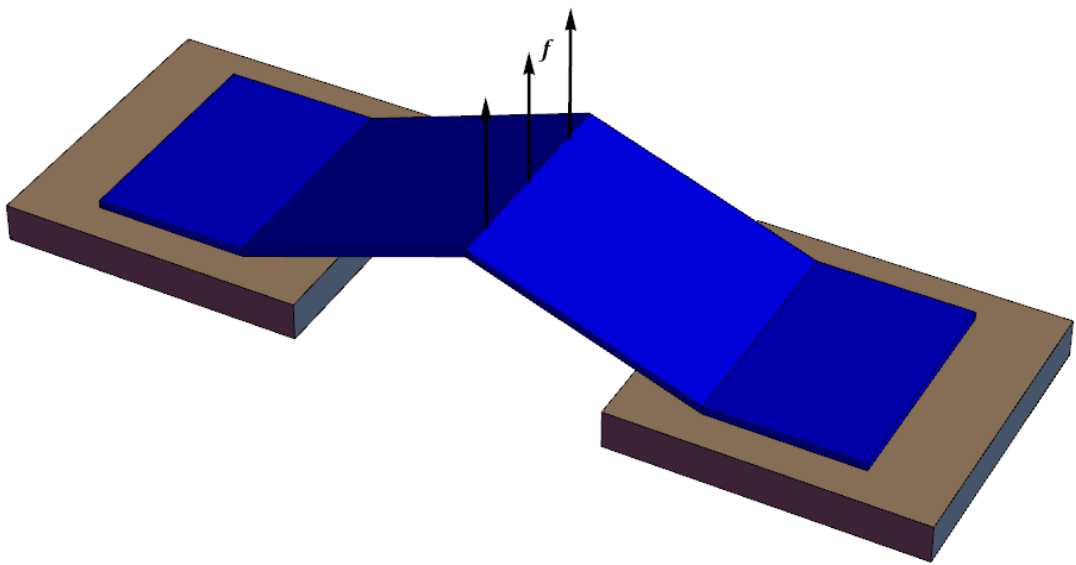

*Figure 13 Schematic of a V-peel test*[33]

We performed V-peel test[33,34] on graphene ribbons adhered to a flat and sinusoidally corrugated substrates (just like the ones described in previous section). The term V-peel test is used by Wan[33] et al in their paper owing to the inverted V-shape assumed by the membrane being peeled as shown in Fig. 13. It is a simple experiment used to determine the adhesion energy wherein a line load or a displacement boundary condition is applied at the middle of a membrane to peel it from the substrate while the edges are fixed. The adhesion energy is obtained from the applied force, measured crack length and peel angle. The goal here is to understand the mechanics of peeling of the graphene membranes at the atomistic scale.

The simulation setup is similar to the one shown in Fig. 2 except here the periodic boundary condition in the width direction is no longer used as we are working with graphene ribbons of finite width now. We took advantage of the symmetry of the peel test setup and simulated only half the membrane. Also, in these simulations the adhesion energy is set at about 0.4 J/m² by adding an LJ potential with $\epsilon \approx 1$ meV and $\sigma = 3.4$ Å to the interactions between the substrate and graphene atoms in addition to the LJ potential from the AIREBO potential. We will first describe the simulations with flat substrates along with a simple analysis to explain the simulation results. Then, we move on to the more complicated peeling simulations with corrugated substrates and use the theoretical approach developed in the previous section to describe the results.

A. Flat Substrates – Theory and simulations

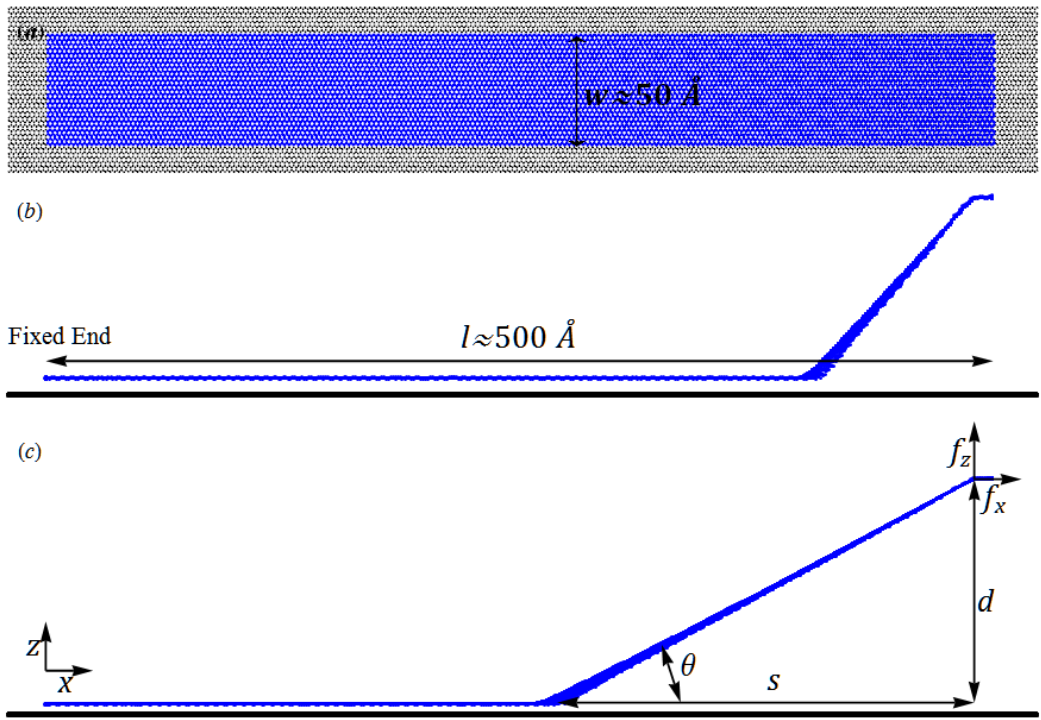

Figure 14 (Color Online) (a) Top view of the initial configuration with substrate and graphene atoms in black and blue colors respectively. (b,c) The self-similar equilibrium configurations at two different specified 'd'.

Peel tests at the macro-scale are conventionally performed with flat substrates with different kinds of boundary conditions and linear or non-linear continuum mechanics analyses exist for each case.[34] Non-linearities usually arise from either large deformations or material models. Here we use one such variant where we apply a displacement boundary condition on one edge, while the other edge is kept fixed. The edge on which the displacement boundary condition is applied is displaced only in the $z$ direction and is held fixed in the $x$ direction. At a given specific displacement $d$, the system is allowed to relax to a minimum energy state as shown in Figs. 14(b) and (c). This is repeated several times, with an increasing $d$ each instance. As $d$ is increased gradually, the membrane is peeled away from the substrate forming a "crack" and simultaneously it is stretched. The resultant force on the displaced edge, $f$ and its components, $f_x$, $f_y$ and $f_z$ are recorded. Also the crack length ($s$) and peel angle ($\theta$) are extracted from the simulation results. The length of the membrane used is about 500 Å and the width is about 50 Å. The free edge is displaced by 175 Å in steps of 0.1 Å.

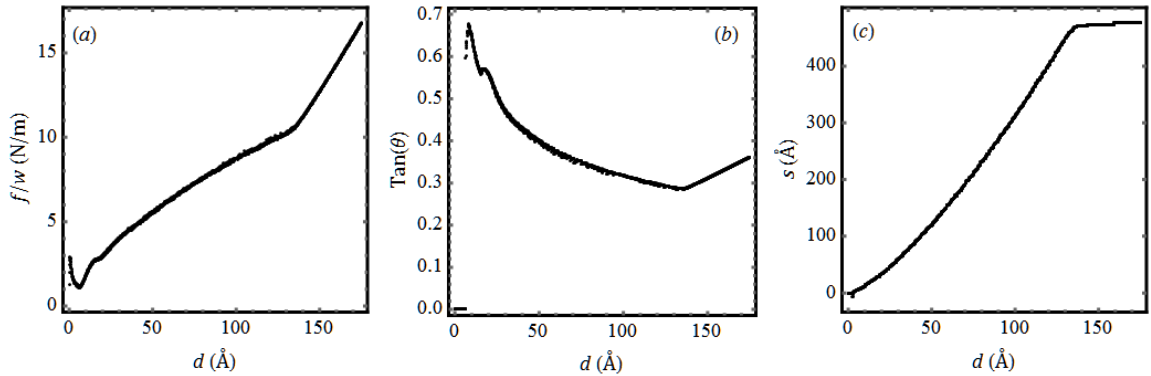

*Figure 15 (a) Total force per unit width, $f/w$ vs Displacement, $d$. (b) Angle, $Tan(\theta)$ vs $d$. (c) Crack length, $s$ vs $d$.*

The results of the simulation are plotted in Figs. 15(a), (b), and (c). The resultant force, $f$, plotted in Fig. 15(a), increases gradually as more of the membrane is peeled from the substrate. At about $d = 137$ Å, the results look different due to the fact that the crack has reached the fixed end as evident from the plot of the crack length (Fig. 15(c)). Here the membrane is only uniaxially stretched and as there is no peeling involved, we are not interested in this part of the results. From continuum theory of the V-peel test,[34] it is known that at equilibrium when the crack is propagating in a self-similar fashion:

$$G = \gamma_0 = \frac{f_z(1 - \cos[\theta] + \frac{\epsilon}{2})}{w \sin[\theta]}. \tag{12}$$

Here G is the energy release rate, $\gamma_0$ is the adhesion energy, $w$ is the width of the membrane and $\epsilon$ is the strain in the delaminated membrane. However when we look at the strain field in the membrane as a function of the $x$ coordinate as obtained from the simulation at $d = 80$ Å (see Fig. 16), we notice that the strain in the membrane is almost uniform. This is due to transmission of the membrane stress through the adhered region of the membrane too, in contrast to the normal peel test at macro-scale. This is possible due to the ability of the atoms in the adhered region to slide over the substrate atoms, which is not the case at macro-scale. As a result of this, the strain energy does not contribute to the energy release rate. Putting $\epsilon = 0$, we calculated the energy release rate using the values of $f_z$ and $\theta$ and the plot is shown in Fig. 17. We can see that the value reaches 0.4 J/m² (indicated by red dashed line in Fig. 17) at about $d = 10$ Å, before which the eq. (12) is not valid as self-similarity is not established yet. It is to be noted that with $\epsilon = 0$, the expression in eq. (12) is now equivalent to the energy release rate in peeling of an inextensible membrane.

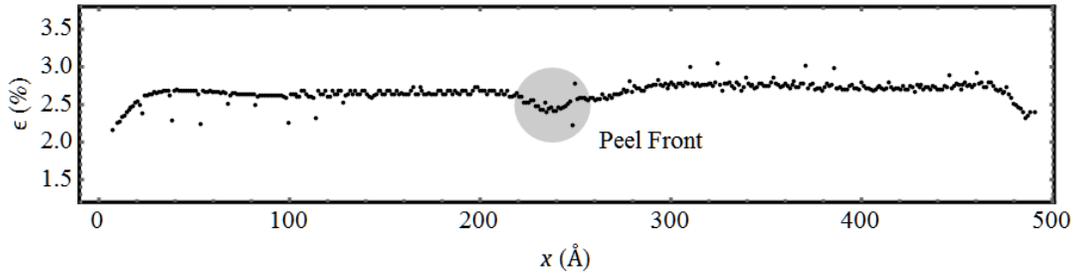

Figure 16 Strain field along the x coordinate in the membrane when $d = 80$ Å.

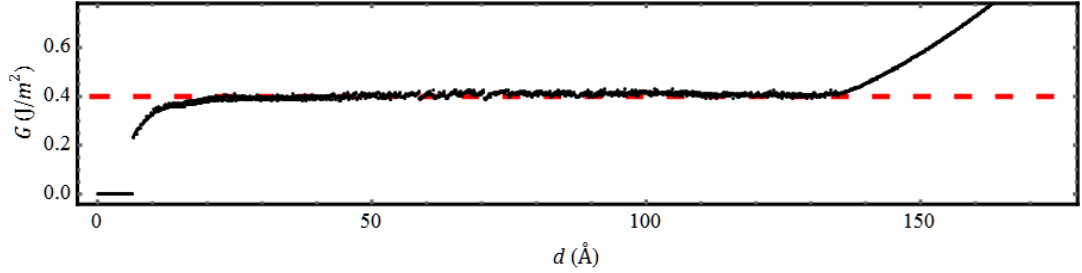

*Figure 17 (Color Online) Energy release rate calculated using eq. (12) putting $\epsilon = 0$.*

Hence, in conclusion, we performed peel test simulation with a graphene ribbon on a flat substrate. The simulation results are found to agree well with the peel test analysis of an inextensible membrane. The reason for this is that graphene ribbon as it is peeled from the substrate slides on the substrate which in turn distributes strain energy uniformly across the delaminated and adhered portions of the membrane. This means that as the membrane is peeled, the strain energy does not contribute to the energy released. In spite of the simulation involving atomistic sliding, the continuum mechanics description holds up quite well.

B. Corrugated Substrates – Theory and Simulations

We now move onto simulations of V-peel tests of graphene ribbons on sinusoidally corrugated substrates to understand the peeling mechanics of graphene at nano-scale. The initial set-up is as shown in Fig. 18(a): a graphene ribbon on a sinusoidally corrugated substrate with amplitude, $c$ and wavelength, $\lambda$. As we learned in the previous sections, graphene will follow the substrate surface profile by achieving a balance between the adhesion energy and bending strain energy. Let the undeformed length of the graphene ribbon be denoted by $l$ and the projected length of the ribbon in its equilibrium configuration be $x_0$ as shown in Fig. 18(a). As in the flat substrate case, one edge is fixed and the free edge is displaced vertically (Fig. 18(b)). At any given displacement, the equilibrium configuration is obtained by energy minimization. As shown in Fig. 18(b), we found that in the equilibrium configuration a portion of the membrane is delaminated while the rest of the membrane still adheres to the substrate. As before, the delamination length, $s$ (or the adhered length, $x = x_0 - s$), the delamination angle, $\theta$, and the peeling force, $f$ are recorded during the simulations. During the course of the simulations, we observed that the graphene atoms slid on the substrate just as in the flat substrate case. This caused change in the conformity of the graphene membrane in the adhered region as the free edge is displaced. So, we also recorded how the amplitude of the adhered region, $a$ changes as the displacement is increased.

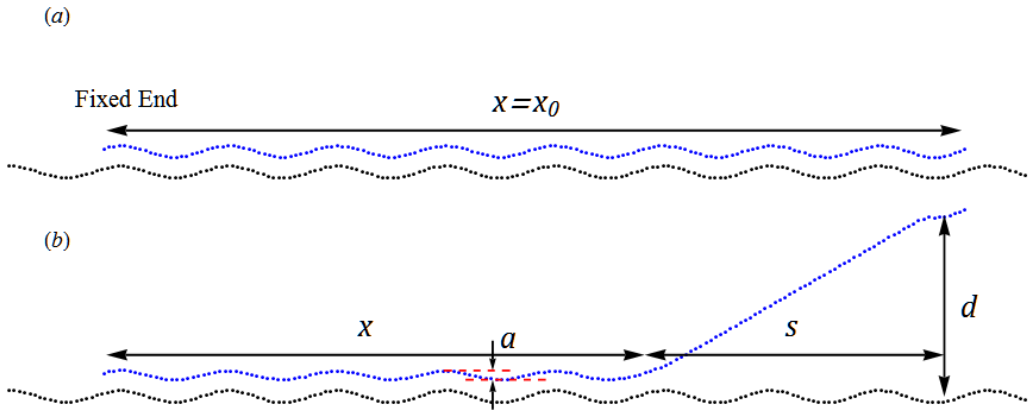

*Figure 18 V-peel test on corrugated substrates. (a) Initial configuration, (b) Equilibrium configuration at a specific 'd'.*

Before presenting the details of the simulation results, let us look at how we can analytically model this problem. The displacement applied at the free edge induces stretching in the whole of the membrane which we assumed to be uniform. We will later verify this assumption using the simulation results. This strain, $\epsilon$ can then be calculated from the constraint:

$$(l(1 + \epsilon) - l_a(a,x))^2 = (d - a \sin[qx])^2 + s^2. \tag{13}$$

Here, $l$ is the initial undeformed length of the graphene membrane and $l_a(a,x)$ is the arc length of the membrane attached to the substrate which can be easily obtained given the sinusoidal shape assumption. This constraint comes from the fact that the free end of membrane is simply displaced vertically upwards. The resultant force needed to displace the free end can be determined from the strain as, $f = Etw\epsilon$.

Given the known variables, substrate amplitude and wavelength ($c$ and $\lambda$), displacement ($d$), the unknowns in this problem are force ($f$), strain ($\epsilon$), angle ($\theta$), adhered length ($x$), adhered region amplitude and equilibrium separation ($a$ and $h$). Assuming $h = h_0$ i.e. fixing the equilibrium separation ($h$) to be the same as that of a flat substrate ($h_0$), the only independent variables here are $a$ and $x$. The rest can be obtained from these two variables: $\epsilon$ and hence $f$ from eq. (13), $\theta$ simply from $d$ and $s$. Hence, the free energy of the system (graphene ribbon, substrate and the adhesive interface) per unit width can then be written as a function of $a$ and $x$:

$$\bar{F}(a,x) = \bar{F}_{adh} + \bar{F}_{ben} + \bar{F}_{str}. \tag{14}$$

Here, $\bar{F}_{adh}$ is the contribution of the adhesive interactions. Using the approach as shown in eq. (7):

$$\bar{F}_{adh}(a,x) = V_f(h_0)x + \sum_{i=1}^{n} \frac{d^i V_f(h)}{dh^i}\Big|_{h=h_0} \frac{(a-c)^i}{i!} \int_0^l \mathrm{Sin}[qx]^i \, dx \tag{15}$$

$$= -\gamma_0 \left( x - \frac{10}{h_0^2}(a-c)^2 \left( x - \frac{\mathrm{Sin}(2qx)}{2q} \right) + \mathcal{O}((a-c)^3) \right).$$

Unlike in eq. (8), as mentioned already, here we fixed the equilibrium separation, $h$ at $h_0$ to simplify the calculations. Also, the interaction of the atoms near the interface of the adhered and detached regions is ignored here. The bending strain energy contribution, $\bar{F}_{ben}$ is given by:

$$\bar{F}_{ben}(a,x) = \frac{D}{2}\int_0^l \left(\frac{d^2g(x)}{dx^2}\right)^2 dx \qquad (16)$$
$$= \frac{D}{4}q^4 a^2 \left(x - \frac{\mathrm{Sin}(2qx)}{2q}\right).$$

Here $g(x) = h_0 + a\,\mathrm{Sin}[qx]$ and any bending strain energy contribution from the region where the membrane goes from adhered to detached is ignored. The strain energy contribution due to stretching induced by the displacement of the free edge, $\bar{F}_{str}$ is then:

$$\bar{F}_{str}(a,x) = \frac{Et}{2}\int_0^l \left(\epsilon + \frac{1}{2}\left(\frac{dg(x)}{dx}\right)^2\right)^2 dx$$
$$= \frac{Et}{2}\left(\begin{array}{c} l\epsilon^2 + \frac{1}{2}\epsilon q^2 a^2 \left(x + \frac{\mathrm{Sin}(2qx)}{2q}\right) \\ + \frac{1}{16}q^4 a^4 \left(\frac{3x}{2} + \frac{\mathrm{Sin}(2qx)}{q} + \frac{\mathrm{Sin}(4qx)}{8q}\right)\end{array}\right). \qquad (17)$$

The equilibrium configuration is then given by minimizing the free energy with respect to the unknowns $a$ and $x$:

$$\frac{\partial \bar{F}}{\partial a} = \frac{\partial \bar{F}}{\partial x} = 0. \qquad (18)$$

Due to the algebraic complexity of the free energy expression, we solved these equations numerically for a given set of parameters. It is to be noticed if the substrate amplitude, $c$ is made zero, then $a$ also goes to zero. This simplifies the free energy to that of membrane adhered to a flat substrate and it can be shown that one can recover the result in eq. (12) (see Appendix C).

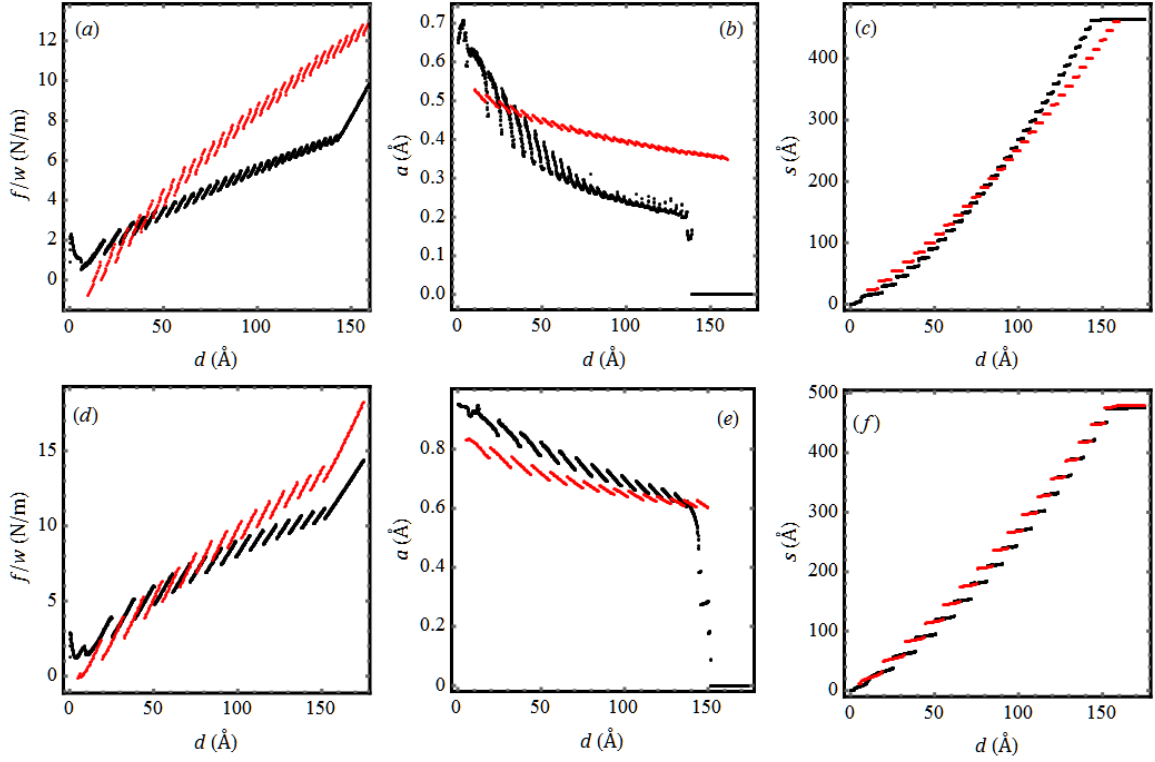

*Figure 19 (Color Online) (a,d) Force per unit length, ($f/w$) vs Displacement, $d$, (b,e) Amplitude, $a$ vs $d$ and (c,f) Crack length, $s$ vs $d$ for $\lambda$ = 15 Å and 30 Å respectively. The data in black and red are from the simulations and theory respectively.*

We now compare the results of the simulations with those from our analysis. Because we assumed that the equilibrium separation does not change from the flat substrate case, we limited our simulations to substrate amplitude of $c = 1$ Å where $h \approx h_0$. The results from the simulations along with results of our analysis are plotted in Fig. 19 with $\lambda = 15$ Å and $\lambda = 30$ Å. It can be noticed that the overall mechanics is discontinuous due to 'instabilities'. We learned from the simulations that these 'instabilities' are due to combined sliding and delamination of the graphene ribbon from the substrate. As the free edge displacement, $d$ is increased initially the membrane just slides resulting in a decrease of the amplitude of the adhered region, $a$ without any change in $x$, the length of the adhered region. Also, we noticed that while sliding, the membrane is pinned to a peak on the substrate. As $d$ is increased further, the membrane 'snaps' by getting detached by a magnitude equal to about half the wavelength, $\lambda$. This snap seems to create slack which gets redistributed into the adhered region increasing the amplitude, $a$ though not back to the initial value. After this snap-off, we noticed that the membrane is pinned at the next available peak on the substrate and now starts to slide again upon increasing $d$. This behavior continues on until the fixed end is reached. The pinning of the membrane at a peak is evident from the nearly discrete increment of the delaminated length as shown in Figs. 19(c) and (f).

If not for the undulating behavior in the overall mechanics as shown in Fig. 19, it is similar to that of peeling from the flat substrate shown in Fig. 15. The force required to delaminate and displace the free end, $f$ increases with increasing displacement, $d$ (Figs. 19(a) and (d)). The delaminated length, $s$ (or equivalently adhered length, $x$) also increases with $d$ (Figs. 19(c) and (f)). On the other hand, the amplitude of the graphene ribbon in the adhered region, $a$ decreases gradually with increasing $d$ (Figs. 19(b) and (e)). The results from our analysis do poorly with $\lambda = 15$ Å case and better with $\lambda = 30$ Å when compared to simulation results. The reason might be the assumption $h = h_0$. Also, it is known graphene exhibits non-linear material properties beyond 1% strain[35] which we definitely surpass in these simulations. In contrast, in our analysis we assumed a constant value $Et$. In spite of its inaccuracy, our analysis captures the essential features of the mechanics involved in this problem quite well.

We asserted earlier that the strain in the membrane is uniform while developing our theoretical analysis. The strain fields at a displacement of $d = 75$ Å are plotted in Figs. 20(a) and (b) for $\lambda = 15$ Å and 30 Å cases respectively. We can clearly see that the strains are quite uniform, hence validating our assertion.

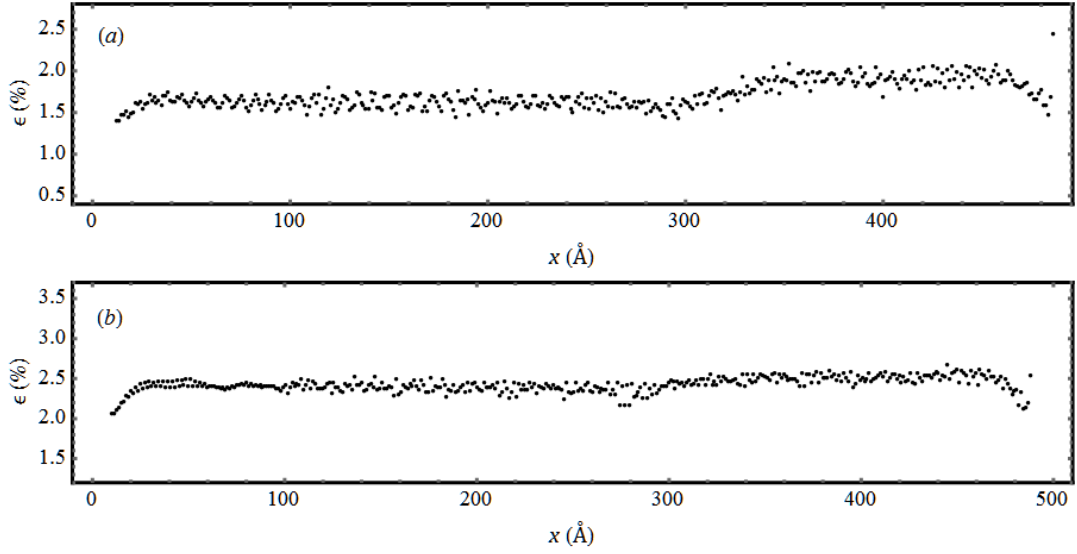

Figure 20 Strain field for (a) $\lambda = 15$ Å case, (b) $\lambda = 30$ Å case with $d = 75$ Å

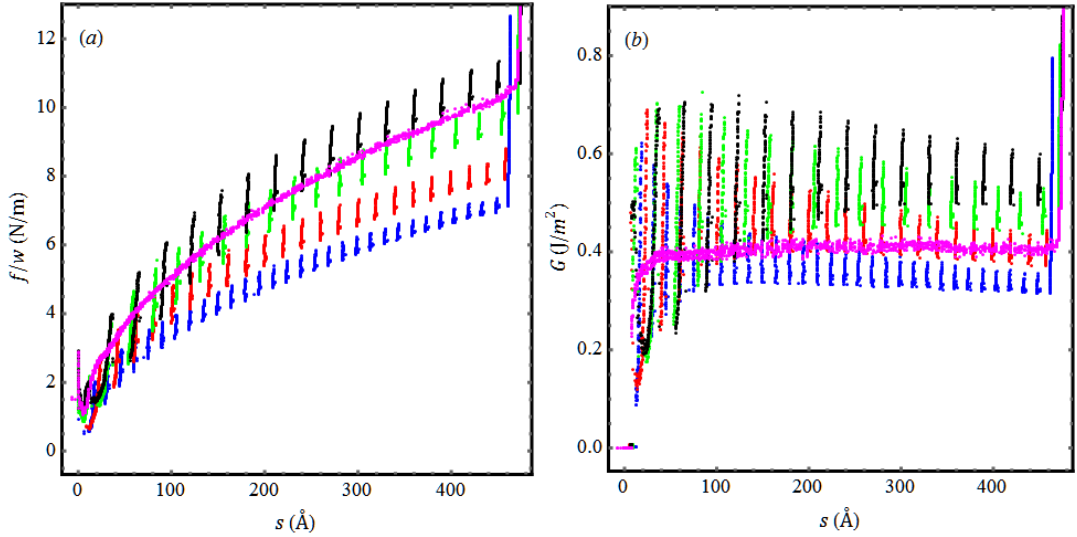

Figure 21 (Color Online) (a) The total force per unit width, $f/w$ and (b) the energy release rate, $G$ according to eq. (12) plotted against the crack length, $s$ for $\lambda = 15$ Å (blue), $\lambda = 20$ Å (red), $\lambda = 25$ Å (green), $\lambda = 30$ Å (black) and flat substrate (magenta).

We plotted a comparison of the magnitude of the force per unit width with respect to the crack length in Fig. 21(a) with different wavelengths along with the limiting case of a flat substrate. Expectedly we see that larger the wavelength, the closer the result to the flat substrate case. Notice that from the periodic nature of these plots, we can easily infer the number of peaks on the substrate and the wavelength. In Fig. 21(b), we plotted the energy release rate using the expression that we used for a flat substrate in eq. (12). The energy release rates for corrugated substrate give undulating values not revealing any direct information about the true adhesion energy as in the flat substrate case. These results are similar to the case of a flat substrate with periodically varying adhesion energy.[36] However, in our case the amplitude of the periodic variation of adhesion strength is coupled to the amplitude of the graphene membrane (see eq. (15)) which in turn depends on the strain in the system in a non-linear manner (see eq. (13)). Thus even as the energy release rate shows a periodic pattern, the amplitude varies in a non-linear intractable manner making it very difficult to extract the adhesion energy from the energy release rate plots.

## IV. SUMMARY

In this paper, we attempt to understand the nano-scale mechanics involved in the adhesion and peeling of graphene membranes. In the first part, we described molecular statics simulations and a companion theoretical analysis where the equilibrium configurations of graphene membranes on sinusoidally corrugated substrates is determined. We learnt through these simulations that the adhesion energy depends on the amplitude and wavelength of the substrate corrugations with larger amplitudes and smaller wavelengths leading to poor conformity. We confirmed a snap-through phenomenon associated with the conformity of graphene that has been observed by several others in the literature. We showed that our analysis compares quite well with the simulation results with both one and two dimensional sinusoidal corrugations and quite accurate when the slopes of the surfaces involved are small. With a specific example, we also showed that monolayer graphene membranes adhere better than multi-layer graphene on corrugated substrates and that the individual layers in multi-layer graphene slide over each other. We found that the ability of the layers to slide coupled with the limited range of LJ interaction potential leads to each layer having a gradually decreasing amplitude from the bottom layer closest to the substrate to the top where the layer can be flat with a sufficiently thick graphene.

We extended our analytical approach to substrates with arbitrary profiles as long as they can be represented by a truncated Fourier series and demonstrated how it works with a simple specific example. We think that our approach might be helpful in understanding adhesion on such useful periodic surfaces like square wave or triangular wave which can be readily fabricated and can be approximated using a truncated Fourier series.

In the second part, the peel mechanics of graphene ribbons on flat as well as sinusoidally corrugated substrates is studied. We found that the mechanics of peeling of the ribbon on a flat substrate can be described in a manner similar to that of an inextensible membrane owing to the sliding of the graphene sheet on the substrate. The mechanics of peeling on corrugated substrates differs significantly from that on the flat substrate and reveals interesting mechanics. In the latter case, we observed instabilities in the way the graphene membrane delaminates from the substrate. We attempted to explain the observed results with the help of a free energy based analysis. This analysis qualitatively captures the essential features of the mechanics involved.

## ACKNOWLEDGEMENTS

This work was supported by NSF grant CMMI 0900832 and the DARPA Center on Nanoscale Science and Technology for Integrated Micro/Nano-Electromechanical Transducers (iMINT). This work utilized the Janus supercomputer, which is supported by the National Science Foundation (award number CNS-0821794) and the University of Colorado Boulder. The Janus supercomputer is a joint effort of the University of Colorado Boulder, the University of Colorado Denver and the National Center for Atmospheric Research.

## APPENDIX A: Comparison with results from the literature

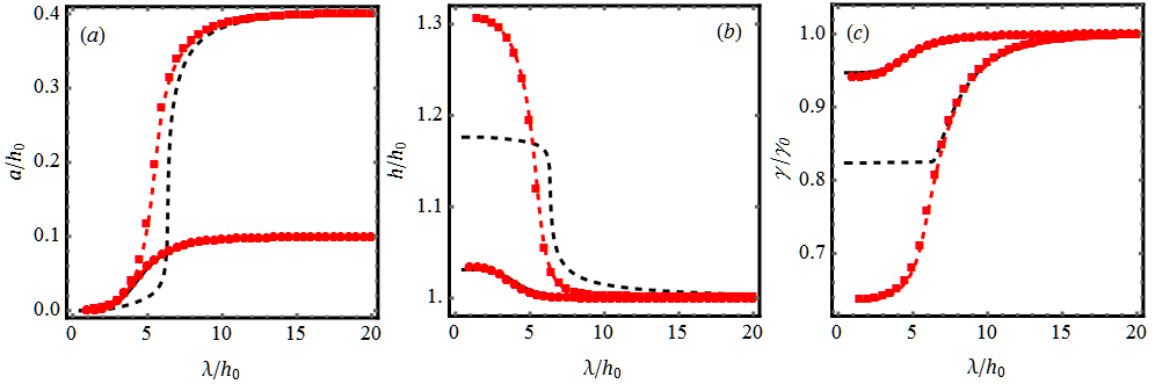

*Figure A1 (Color Online) Plots comparing our calculations with those of Aitken and Huang: non-dimensional (a) Amplitude, (b) Mean separation and (c) Adhesion energy are plotted against the non-dimensional wavelength. The red curves (from optimization of eq. (8)) and the circle/square symbols (from direct integration; see supplementary text) are our results while the black curves are the results of Aitken and Huang. Here the solid curves and circular symbols are calculations done with $c/h_0 = 0.1$; the dashed curves and square symbols are calculations done with $c/h_0 = 0.4$.*

We compared the results from our calculations with those of Aitken and Huang[22] where they do not use the Derjaguin approximation but approximately calculate the adhesion energy from the non-local $V_{pot}$ for sinusoidal surfaces. We used the same potential and parameters that were used in their paper for the purpose of this comparison. The potential they used is different from $V_f$ in eq. (3); it accounts for interaction between a surface of atoms with a semi-infinite body. It is straightforward to replace $V_f$ with the potential they used. The parameters used are $\frac{D}{\gamma_0 h_0^2} = 6.94$ and $\frac{c}{h_0} = 0.1$ or $0.4$.

The results are shown in Fig. *Figure A1*. The plots from left to right show the non-dimensional amplitude, mean separation and the adhesion energy as a function of the wavelength of the substrate. The membrane conforms to the substrate very closely at higher wavelengths while it is relatively flat at lower wavelengths. It can be seen that there is a good agreement in general between the three methods shown here – Aitken and Huang's (black curves), our method with $\hat{F}_{adh}$ calculated by direct integration (see supplementary text for details) (circle and square symbols) and our method where we use the expression in eq. (7) with 40 terms to calculate $\hat{F}_{adh}$. For the lower amplitude ($c/h_0 = 0.1$, solid curves and circular symbols), the three methods give exactly the same result; while for the higher amplitude ($c/h_0 = 0.4$, dashed curves and square symbols), though our two approaches still agree quite well, our results differ considerably from Aitken and Huang's results. At the higher amplitude, Aitken and Huang's calculations underestimate (overestimate) the mean separation (adhesion energy) compared to our calculations even though the amplitude predicted is quite similar. This might be attributed to the approximations used by Aitken and Huang ($c \ll h_0$) which limits the use of their method at high amplitudes ($c < 0.5 h_0$) or to the Derjaguin approximation we used.

We also compared our analysis results with the simulation results of Li and Zhang[27] where they numerically integrate the free energy and minimize it for substrates with 1d sinusoidal shapes. Using the same potential and parameters as they used, the results of the comparison are presented in Fig. Figure A2 where the ratio of amplitudes of the membrane and the substrate ($a/c$) is plotted against the ratio of bending rigidity to the depth of the LJ potential well ($D/\epsilon$) at four different values of substrate wavelength to amplitude ratios ($\lambda/c$). As can be seen, our results do well at higher wavelength to amplitude ratios i.e. when the surface is relatively flatter consistent with limitations of Derjaguin approximation.

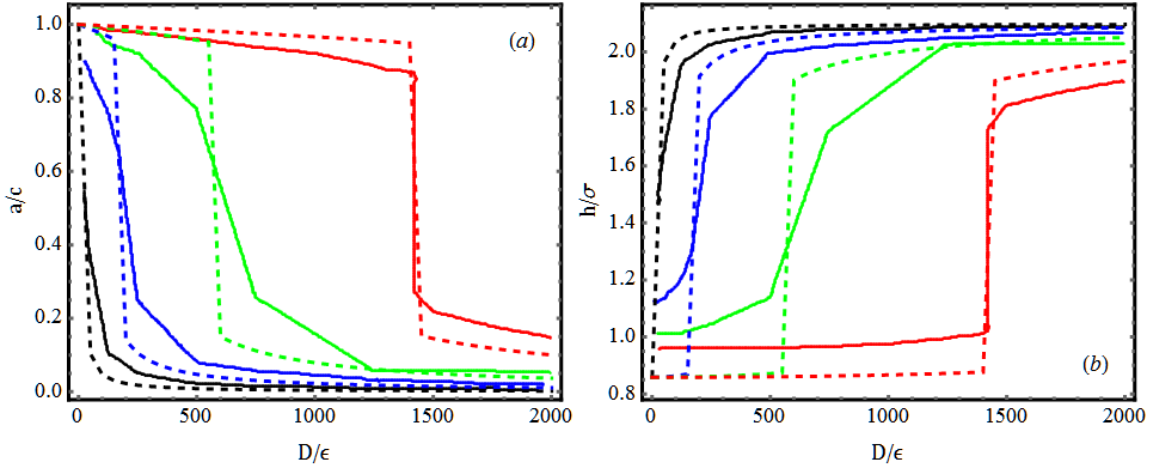

*Figure A2 (Color Online) Plots comparing our calculations with those of Li and Zhang: non-dimensional (a) Membrane amplitude and (b) Mean separation are plotted against the non-dimensional bending rigidity. The solid curves are Li and Zhang's results[27] and the dashed curves are our results. The black curves are for $\frac{\lambda}{c} = 4$, blue curves – 6, cyan curves – 8 and red curves – 10.*

We then compared our results with 2d sinusoidal surfaces ($s(x) = c \, \text{Sin}[qx] \, \text{Sin}[qy]$) with those from Li and Zhang's simulations.[28] Our results agree quite well with their results as shown in Fig. Figure A3 where again $a/c$ is plotted against varying $D/\epsilon$ at different wavelengths.

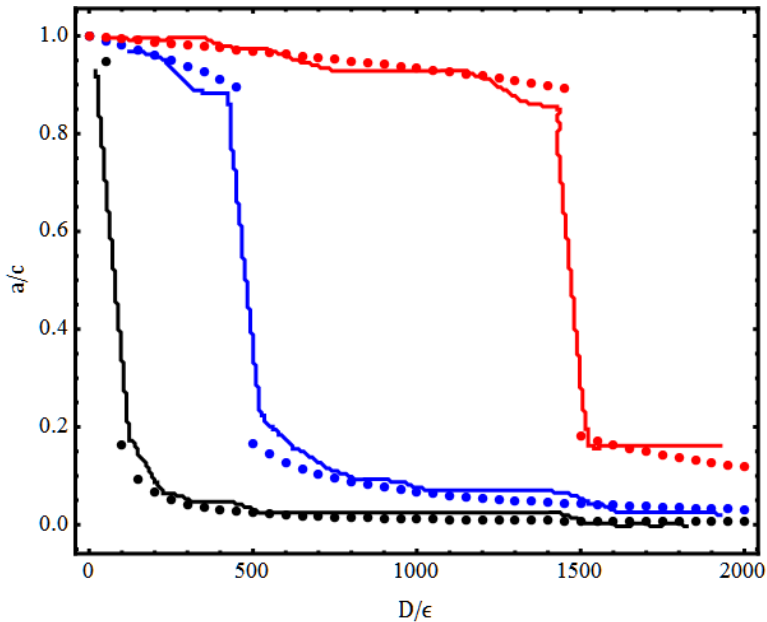

*Figure A3 (Color Online) Plots comparing our calculations with those of Li and Zhang for 2d sinusoidal surfaces. The non-dimensional amplitude is plotted as the non-dimensional bending rigidity is varied from 0 to 2000. The points are our results while the solid curves are from Li and Zhang.[28] The black curve is for a wavelength of 40 Å, blue is for 60 Å and red is for 80 Å.*

APPENDIX B: Generalized free energy using 2D complex Fourier series

Let the substrate be described by the truncated complex Fourier series,

$$s(x,y) = \sum_{m,n} c_{mn} e^{i(q_m x + q_n y)}. \tag{B1}$$

$(m, n \in \mathbb{Z}, -M \leq m \leq m, -N \leq n \leq N, \ 0 \leq x, y \leq L, c_{00} = 0, c_{mn} = c^*_{-m-n}, q_m = 2\pi m/L)$

and let us assume that the graphene membrane follows the curve,

$$g(x,y) = h + z(x,y) = h + \sum_{m,n} a_{mn} e^{i(q_m x + q_n y)}. \tag{B2}$$

$(a_{mn} = a^*_{-m-n}, a_{00} = 0)$. In this case, the free energy per unit volume following our approach will be:

$$\hat{F}([a_{mn}], h) = \frac{D}{2} \sum_{m,n} |a_{mn}|^2 (q_m^2 + q_n^2)^2 + V_f(h) \\ + \sum_{j=1}^{p} \frac{d^j V_f(h)}{dh^j} \frac{1}{j!} \sum_{\substack{\sum_{m,n} l_{mn}=j \\ \sum_{m,n}(m,n)l_{mn}=0}} \frac{j!}{\prod_{m,n} l_{mn}!} \prod_{m,n} (a_{mn} - c_{mn})^{l_{mn}}. \tag{B3}$$

Here, $c_{mn}$ and $a_{mn}$ are the Fourier coefficients and are complex numbers; $h$ (equilibrium separation) and $a_{mn}$ being the unknowns. The internal summation in the nested summation of the last term is a result of a multinomial expansion where $l_{mn}$ are the exponents which have to obey the constraints $\sum_{m,n} l_{mn} = j$, $\sum_{m,n} m l_{mn} = 0$ and $\sum_{m,n} n l_{mn} = 0$. The latter two of the three constraints come from the non-zero terms after integration of each term in the multinomial expansion.

# APPENDIX C: Peeling from corrugated substrates – Limiting case

The limiting case for peeling from corrugated substrates would be a flat substrate where $c = 0$. In the absence of corrugations, the membrane should also be flat i.e. $a = 0$. Hence, the free energy as described in section III-B (see eqs. (14)-(17)) can now be written as:

$$\bar{F}(x) = -\gamma_0 x + \frac{Et}{2} l\epsilon(x)^2. \tag{C1}$$

Here, $\epsilon = \frac{(d^2+s^2)^{\frac{1}{2}}-s}{l}$ and $s = l - x$ with all the symbols retaining their original meanings in III-B. Hence minimizing the free energy with respect to $x$ leads us to:

$$\begin{aligned}\frac{d\bar{F}(x)}{dx} &= -\gamma_0 + Et\epsilon(x)\frac{d\epsilon(x)}{dx} = 0 \\ \Rightarrow \gamma_0 &= Et\epsilon\left(1 - \frac{s}{(d^2+s^2)^{\frac{1}{2}}}\right).\end{aligned} \tag{C2}$$

Notice that $f = Et\epsilon$ and $\cos[\theta] = \frac{s}{(d^2+s^2)^{\frac{1}{2}}}$, hence $\gamma_0 = f(1 - \cos[\theta])$ which is equivalent to the inextensible membrane version of the eq. (12). Thus we recover the energy release rate for the flat substrate as the limiting case for peeling from a corrugated substrate.